\documentclass[twocolumn,letterpaper,pre]{revtex4}
\usepackage{epsfig}
\usepackage{subfigure}

\begin{document}
\title{Selective control of the apoptosis signaling network in heterogeneous cell populations}
\author{D. Calzolari}
\author{G. Paternostro}
\affiliation{Burnham Institute for Medical Research, 10901 North Torrey Pines Road,
La Jolla, California 92037}
\author{P. L. Harrington Jr.}
\author{C. Piermarocchi}
\author{P. M. Duxbury}
\affiliation{Physics and Astronomy Department,
Michigan State University, East Lansing, MI 48824, USA.}
\date{\today}

\begin{abstract}
{\it Background.} Selective control in a population is the ability to control a member of
the population while leaving the other members relatively unaffected. The
concept of selective control is developed using cell death or apoptosis in
heterogeneous cell populations as an example. Control of apoptosis is
essential in a variety of therapeutic environments including cancer where
cancer cell death is a desired outcome
and Alzheimer's disease where neuron survival is the desired
outcome.   However in both cases these
responses must occur with minimal response in other cells exposed to treatment
that is, the response must be selective.

{\it Methodology and Principal findings.}  Apoptosis signaling in
heterogeneous cells is described by an ensemble of gene networks with identical
topology but different link strengths. Selective control depends on the
statistics of signaling in the ensemble of networks and we analyse the effects
of superposition, non-linearity and feedback on these statistics. Parallel
pathways promote normal statistics while series pathways promote skew
distributions which in the most extreme cases become log-normal. We also show
that feedback and non-linearity can produce bimodal signaling statistics, as
can discreteness and non-linearity. Two methods for optimizing selective
control are presented. The first is an exhaustive search method and the second
is a linear programming based approach. Though control of a single gene in the
signaling network yields little selectivity, control of a few genes typically
yields higher levels of selectivity. The statistics of gene combinations
susceptible to selective control in heterogeneous apoptosis networks is studied
and is used to identify general control strategies. 

{\it Conclusions and Significance.}  We have explored two methods
for the study of selectivity in cell populations.
 The first is an exhaustive search method limited
to three node perturbations. The second is an effective linear model, based on
interpolation of single node sensitivity, in which the selective combinations
can be found by linear programming optimization. 
We found that selectivity is promoted by acting on the least
sensitive nodes in the case of weak populations, while selective control of
robust populations is optimized through perturbations of more sensitive
nodes. High throughput experiments with heterogeneous cell
lines could be designed in an analogous manner, with the further 
possibility of incorporating the selectivity optimization process
into a closed-loop control system.

\end{abstract}

\date{\today}

\maketitle

\section{Introduction}

Living cells carry out their functions, such as
working, reproducing and dying, by appropriate response to
extracellular and intracellular inputs to a complex network
of signaling pathways.  Genes which code for the proteins
in these pathways are controlled by regulatory proteins
which up-regulate or down-regulate these genes, depending
on inputs to the signaling network.  The enormous effort
currently directed at understanding signaling networks may be
subdivided into two areas, firstly extracting faithful wiring diagrams
for the networks and, secondly developing methods to understand
and control the messages which pass through them.
In this contribution we develop the concept of selective control
in diverse cell populations, and introduce computational methods
which optimize selectivity for a particular signal for a designated 
member of a cell population.

We introduce the concept of selective control or selectivity
 in cell populations as the
requirement of finding a set of inputs which induce one member of the
population to produce a desired response while ensuring that the remaining
members of the population have a minimal response.  In the case of apoptosis or
cell death, which we use as an illustrative example, we consider a population
of cells and seek methods to kill a selected member of the population while
ensuring the survival of the others in the population. Control of apoptosis is
essential in a variety of therapeutic environments including cancer where
cancer cell death is a desired outcome\cite{petak2006,klein2006,
johnstone2002, fitzgerald2006}
and Alzheimer's disease where neuron survival is the desired
outcome\cite{lu2004,pompl2003,ankarcrona2005}.   However in both cases these
responses must occur with minimal response in other cells exposed to treatment
that is, the response must be selective.

Though striking progress is occuring in the extraction
of networks using a range of experimental data\cite{perkins2006,barrett2006},
 knowledge of signaling networks
remains predominantly at the level of topology rather than detailed knowledge
of the rate constants and non-linear message passing which
occurs in the networks.  Models to
distinguish between members of a population of cells,
 for example different cancer cells
and different normal tissue types, require differences in
message passing parameters and/or expression levels of the genes in the
network.  Here, the computational procedures for selectivity
in cell populations are elucidated using heterogeneous
populations, where members of the population
are distinguished by having message-passing efficiencies drawn
from homogeneous random distributions.

Models of message passing in gene networks range from binary
models with discrete message passing rules\cite{li2004,shmulevich2005,dealy2005}
to non-linear ordinary differential equations\cite{aldridge2006} and to stochastic
spatio-temporal models\cite{kholodenko2006} which are simulated using partial differential
equations or Monte Carlo methods.  Questions of interest also vary greatly,
from generic questions about the number of attractors and their stability
in random networks\cite{zhou2005,huang2005,klemm2005,colizza2005, drossel2005}
to modeling the detailed dynamics of gene concentrations
in particular pathways\cite{eissing2005,buchler2005,bagci2006},
and to the cellular response such as control
of flagellar rotation in bacteria responding to chemotaxis.
   Some of the tools developed for the analysis of
metabolic networks, both dynamically and using steady state flux balance
approaches, can be profitably extended to signaling networks\cite{covert2001,lee2006}.
In the flux balance approach (FBA), the ouput of a cell may be optimized with respect to
an objective function and subject to the constraints of flux balance at
each node in the network.

Though there are conceptual similarities between the FBA
and our approach, there are also critical differences.  Firstly as described
in Section II, rather than
using flux balance, we require
message passing rules which describe how a gene responds to the state
of its neighbors.  To illustrate the effects of different rules,
 we use an important example from systems biology, the apoptosis network.
In particular we discuss the statistics of death signals produced by
continuum and discrete message passing rules in this network.

In Section II we also develop the concept of selective
control.  Rather than optimizing the objective
for one cell or metabolic network, as occurs in FBA,
we seek to optimize the response
of one cell in a population while minimizing the response of other
cells in the population.  In this section selective control is
demonstrated using exhaustive search over drug combinations in discrete models and
using an approximate linear programming approach. 
Though drugs affecting specifically every node of the apoptosis network
are not yet available, this a very active field of pharmacological research
and it is probably one of the biological networks where this 
ideal situation, from the control point of view, is closest to reality \cite{reed2005}.
Moreover, the network we use is probably still an incomplete representation of the apoptosis 
network, both for the topology and for the kinetic parameters. Nevertheless, several authors 
have shown that useful results can be obtained from partially characterized 
models of biological networks \cite{duarte2007,albert2003}.
Section III contains a discussion of the main points of the paper.

\section{Results}

\subsection{Statistics of signaling}

We have modified the apoptosis network, hsa04210,
of the Kegg database to that presented in Fig. 1, where
network complexes consisting of several genes are
split into individual nodes.    Recent work
emphasizes the importance of positive feedback between
CASP3 and CASP8 denoted by the long dashed arrow in the
figure\cite{eissing2005,bagci2006}, though the importance of
this feedback is not universally accepted.
\begin{figure*}
{\centering \resizebox*{1.8\columnwidth}{!}
{\rotatebox{0}{\includegraphics{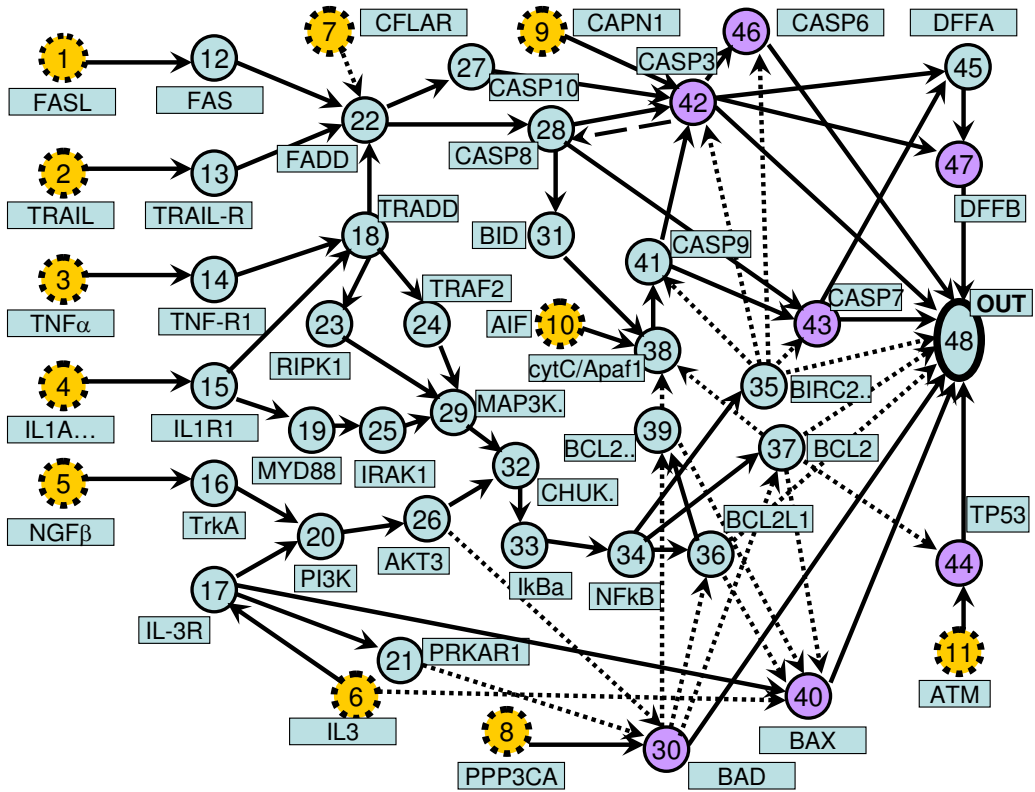}}}\par}
\caption{The human apoptosis network modified, as described in the text,
from that in the Kegg database (hsa04210).
There are 47 genes in the figure and an additional node which
we label the output node.  The 47 genes may be
catagorized (see Table 1) as: input genes
(dashed circles); membrane genes; death genes;
life genes; and finally the remaining genes in the network.  The output,
or "death" node (48) is added to represent the cumulative effect of many
genes implicated in the onset of the cell death machinery.  In this figure, 
solid lines indicate promotion while short dashed links indicate inhibition.
The long dashed link between CASP3 and CASP8 adds the possibility of feedback 
in the apoptosome.}
\end{figure*}
There are 47 genes in Figure 1 and an additional node which
we label the output node.  The 47 genes may be
catagorized (see Table 1) as: input genes
(dashed circles in the Fig. 1), which transmit
signals to the network from the other parts of the
cellular network; membrane genes which code for membrane
proteins and complexes bound to membrane proteins; death genes which
signal onset and execution of apoptosis; life genes which
reduce apoptotic signals and are upregulated in many cancer
cells; and finally the remaining genes in the network.  The output,
or "death" node (48 in Fig. 1) is added to represent the cumulative effect of many
genes implicated in the onset of the cell death machinery.
\begin{table}
\caption{Typical roles of genes in the signaling network.}
\hspace{0.2in}
\begin{tabular}{|r|c|c|c|l|}
\hline
{\bf Input}  & {\bf life} &   {\bf death}    & {\bf membrane}   & {\bf other} \\

   FASL         &     BIRC2..     &   BAD            &  FAS        & CASP10 \\
   TRAIL         &    BCL2L1      &   CASP6         &   FADD       & CASP9\\
   TNF-$\alpha$  &   BCL2       & CASP3      &  TRADD        & CASP8 \\
   IL1A         &    BCL2..      &  CASP7             & TNF-R1         & BID\\
   NGF$\beta$   &          &  BAX             &  RIPK1        & CytC/Apaf1\\
   IL3      &          &      DFFB      & TRAF2        &  MAP3K\\
   CFLAR     &          &     AIF       &   TRAIL-R       & PI3K\\
   PPP3CA        &          & TP53           &  IL1R1        & AKT3\\
   CAPN1     &          &               &  MYD88        &  CHUK..\\
   AIF    &          &               &   IRAK1       & IkBa\\
   ATM    &          &               &    NGF$\beta$       & NFkB\\
    &          &               &    TrkA     & PRKAR1\\
     &          &               &    IL3     & DFFA\\
    &          &               &    IL-3R      & \\
\hline
\end{tabular}
\end{table}
We consider two types of models,
those where the gene activities, $a_i$,
are continuous and those with discrete gene activities, $m_i$.
  We use models with continuous activities to illustrate the
generic statistics of signal propagation in the apoptosis network,
while discrete models are more convenient for the exhaustive search
methods used in the selectivity
studies discussed in Section III.
In the discrete models the gene activity has discrete values
up to a maximum value $M$, so that $m_l = 0,1...,M$.
Binary networks, where $M=1$ have received the most attention,
following the work of Kauffman \cite{kauffman1993}.  We considered three discrete cases,
$M=1,2,10$, though here we focus upon M=10 which is closer to the non-linear continuous
behavior observed in experiment.

Each gene receives signals from the genes that it
is connected to in the signaling network of Figure
1.  The signal arriving at a gene depends on the strength
of the connections to its neighbors in the network.
We define the strength of these connections
to be $\omega_{ij}$ between the $i^{th}$ and $j^{th}$ genes.
Since the network is directed, $\omega_{ij}\ne \omega_{ji}$.
The values of $\omega_{ij}$ are poorly characterized
even in metabolic networks where they correspond to
reaction rates.  In the absence of detailed
knowledge about these connections we take them to be
random variables and in this way develop a generic understanding
of signal propagation in heterogeneous cell populations.
The link weights, $\omega_{ij}$ have positive
random values for promotion links and negative
random values for inhibition links.

\subsubsection{Continuous models}

In the continuous models, the edge weights $\omega_{ij}$ are
uniform continuous random variables and each gene has activity
$a_i$ which is a continuous variable.  The signal arriving
at a gene is given by the sum,
\begin{equation}
s_l = \sum_{j\epsilon n(l)} w_{jl} a_j,
\end{equation}
where $n(l)$ is the set of genes which send signals to gene $l$.
The signal $s_l$ arriving at gene $l$ may be positive or
negative, where a negative signal implies inhibition.
However, the activity of a gene must be positive or zero
so that a negative signal at a gene implies
complete inhibition and the gene is switched off, so that its
activity is set to zero.  This is a basic
non-linearity in signaling networks.

In addition, gene activity levels are often observed to depend in a
non-linear way on the signals arriving at the gene.
A common approximation
to the nonlinear response in gene activity is the Hill equation\cite{weiss1997},
\begin{equation}
a(s) = {c s^b \over 1 + d s^b},
\end{equation}
where $c/d$ is the saturation value of the gene activity, $d$ determines the
onset of saturation and the exponent $b$ is the cooperativity index.  The
case, $b=1$, is Michaelis-Menten behavior characteristic
of a chemical reaction in the presence of a substrate.
The simplest case $b=1$, $c=1$, $d=0$ is the
majority rule signaling procedure, given by $a_l = s_l$ for
$s_l$ positive and $a_l = 0$ for $s_l$ negative.

There are several procedures for simulating signal
propagation through networks.
In binary networks there has been considerable study of
synchronous as opposed to asynchronous updates,
where in the former case the gene activity levels
at time $t$ are used to update all of the activity
levels at time $t+1$.  In contrast asynchronous methods
update gene activity randomly, for example one randomly
chosen gene at a time, to model stochastic behavior.
The number of attractors found in binary networks
appears to depend on the update procedure\cite{drossel2005}.
In the absence of the feedback link between
nodes 42 and 28 in Fig. 1, the
network presented there has no loops.  In this
case signal propagation through the network is
deterministic and non-chaotic.  Furthermore signal
propagation through the loopless network can be carried out
in one sweep of the network by ordering the nodes
according to their distance from the inputs.  The
nodes are then updated in order of their distance from
the input, a procedure which we denote the optimal signaling
algorithm(OSA).  As described later, this procedure
can be modified to take into account the feedback induced
 by the link between CASP3 and
CASP8 in Fig. 1\cite{eissing2005,bagci2006}.  In the absence
of this link, the longest path from the inputs to the output has 15
links and hence the OSA algorithm is completed in 15 timesteps.

We directly tested the effect of asynchronous, synchronous and
OSA procedures on signaling in the loopless apoptosis network and
found that they produce essentially the same results.  This result is
at first counterintuitive as undirected random networks show
more complex behavior, such as chaos, and larger sets
of attractors.  However directed networks are determininistic
so that for a given set of inputs in a network with fixed edge
weights,  and using deterministic
message passing rules, there is a unique output.
Statistical variations do occur however when the link
weights are varied or stochastic noise is added to the
message passing rules.  Since we are interested in
heterogeneous populations, which is the analog of
quenched disorder, we consider variations in
signaling due to variations in the edge weights.

Typical statistic behavior of signals passing
through the loopless
apoptosis network are presented in Fig. 2 for
an important gene, NFkB, in the interior of the network
and also for the cumulative death signal at the output.  These
distributions are found by simulating
50000 different networks, where
each network has links (values of $\omega_{ij}$) having
weights drawn from a uniform distribution on the interval [0,1].
These simulations were carried out for a model with continuous
activities $a_i$, where Eq. (1) is used
to find the total signal arriving at a gene and we use the linear
relation $a_i=s_i$ for positive signals and $a_i=0$ for negative
signals. The input genes in the network were assigned random values
on the interval $[0,1]$.  Positive values of the
signal arriving at the output node indicate cell death, while
negative values denote life.  Although the output node
statistics (see Fig. 2a) is somewhat skew it is not too
far from a normal distribution,
however the statistics at NFkB (Fig. 2b) is highly skew and is
almost lognormal.  We now provide
a simple explanation for the contrasting statistical
behavior occuring for NFkB (34 in Fig. 1) and the output node (48 in Fig. 1).
\begin{figure}
\mbox{\subfigure[]{
  \epsfig{file=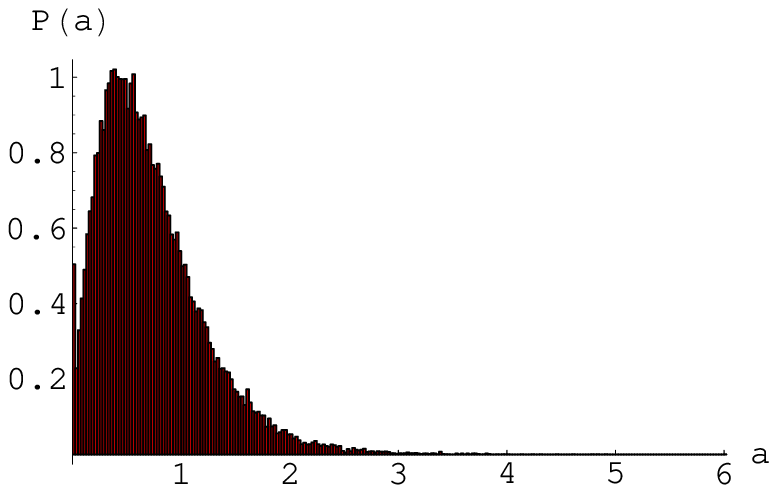 , width=0.8\columnwidth}}}\\
\mbox{\subfigure[]{
  \epsfig{file=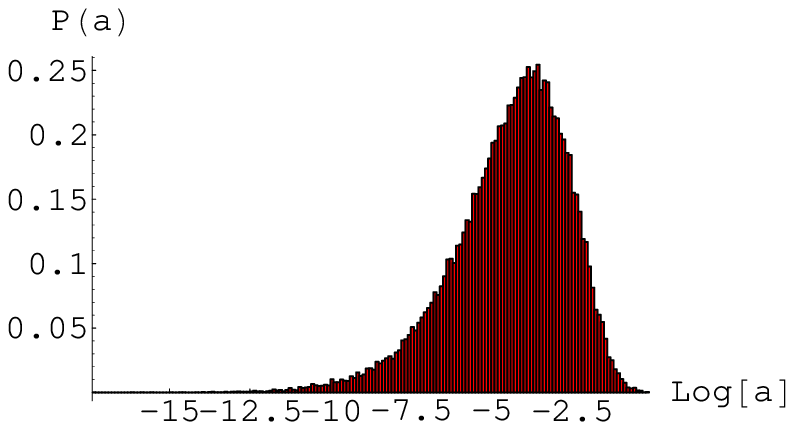 , width=0.8\columnwidth}
}}
\caption{ The distribution of gene activities in heterogeneous cell
populations with a population size of 50,000.  These
distribution are for majority rule signaling with
positive and continuous gene activities. a) The activity
of the output node is close to a normal
distribution due to many pathways arriving at the
output node; b) Signal statistics at the gene NFkB, which
is an internal node lying at the end of a chain
of links (see Fig. 1), is close to log-normal. (see
discussion in the text).  }
\end{figure}

Simplified models elucidating the origin of the
signaling statistics observed in Fig. 2
are presented in Figs. 3 and 4.
Many paths enter the death node (48) and
this is simplified to a set of independent parallel
paths in Fig. 3.  According to Eq. (1), the
death signal is then a sum of random variables and it
is well known in that case that 
the statistics of the signal should be
a normal distribution, in the asymptotic limit.
The observed near normal behavior observed
for the death node is then due to the
fact that many parallel paths enter the death node.
Deviations from the ideal normal distribution
are expected for several reasons, including the fact that
the activities cannot be negative, the presence of correlations
in the signals entering the death node, and due to the
fact that we are far from the asymptotic limit.
\setlength{\unitlength}{0.01in}
\begin{figure}
\begin{picture}(200,100)
\thicklines
\put(-5,53){\circle{6}}
\put(-3,65){\makebox(0,0){$a_1$}}
\put(-3,50){\vector(2,-1){97}}
\put(30,40){\makebox(0,0){$\omega_1$}}
\put(45,53){\circle{6}}
\put(47,65){\makebox(0,0){$a_2$}}
\put(47,50){\vector(1,-1){48}}
\put(70,40){\makebox(0,0){$\omega_2$}}
\put(97,53){\circle{6}}
\put(97,65){\makebox(0,0){$a_3$}}
\put(97,47){\vector(0,-1){44}}
\put(110,40){\makebox(0,0){$\omega_3$}}
\put(150,53){\circle{6}}
\put(147,65){\makebox(0,0){$a_4$}}
\put(147,50){\vector(-1,-1){48}}
\put(150,40){\makebox(0,0){$\omega_4$}}
\put(200,53){\circle{6}}
\put(197,65){\makebox(0,0){$a_{5}$}}
\put(197,50){\vector(-2,-1){97}}
\put(197,40){\makebox(0,0){$\omega_5$}}
\put(97,0){\circle{6}}
\put(120,-5){\makebox(0,0){$a_{out}$}}
\end{picture}
\caption{A parallel combination of signaling pathways with
no series connections. For a large number of parallel
connections, the output activity $a_{out}$
is normally distributed (see text).}
\end{figure}
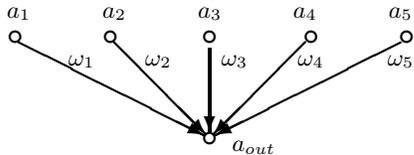

In contrast, NFkB is at the end of a chain of connections
(see Fig. 1) and a simplified model of this connectivity is
 illustrated in Figure 4.  In this case Eq. (1) yields,
\begin{equation}
a_{out} = \prod_{l=1}^n \omega_l a_1~.
\end{equation}
This is a random multiplicative process, so that if
the link variables $\omega$ have random noise, then the
output signal, $a_{out}$ asymptotically obeys log-normal statistics.
The log normal distribution, in the variable $x$, is given by,
\begin{equation}
p(x) = {1\over (2\pi)^{1/2} x\sigma} e^{-(ln(x)-\mu)^2)/2\sigma^2}~,
\end{equation}
which is typically highly skew and exhibits large fluctuations.
Here $\sigma,\ \mu$ are parameters
in the distribution. The statistical variations of signals
arriving at genes in complex networks clearly varies a
great deal depending on the local connectivity of the genes.
In cases where there are mostly linear pathways, as is
believed to occur in some cancer cells, there is
a greater potential for strong fluctuations
in signaling statistics.
\setlength{\unitlength}{0.01in}
\begin{figure}
\begin{picture}(200,75)
\thicklines
\put(-3,25){\circle{6}}
\put(-3,40){\makebox(0,0){$a_1$}}
\put(0,25){\vector(1,0){44}}
\put(25,15){\makebox(0,0){$\omega_1$}}
\put(47,25){\circle{6}}
\put(47,40){\makebox(0,0){$a_2$}}
\put(50,25){\vector(1,0){44}}
\put(75,15){\makebox(0,0){$\omega_2$}}
\put(97,25){\circle{6}}
\put(97,40){\makebox(0,0){$a_3$}}
\put(100,25){\vector(1,0){44}}
\put(125,15){\makebox(0,0){$\omega_3$}}
\put(147,25){\circle{6}}
\put(147,40){\makebox(0,0){$a_4$}}
\put(150,25){\vector(1,0){44}}
\put(175,15){\makebox(0,0){$\omega_4$}}
\put(197,25){\circle{6}}
\put(197,40){\makebox(0,0){$a_{out}$}}
\end{picture}
\caption{A signaling pathway with four steps in series and
with no parallel connections.  If the number of steps
in the pathway is large, the ouput activity $a_{out}$
obeys log-normal statistics (see text).}
\end{figure}
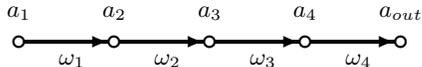

Non-linearity is a hallmark of genetic response and we have
studied the effect of a variety of non-linear activity-signal
behaviors on signaling in the apoptosis network.
We tested the effect of various Hill equation parameters on the
activity statistics of the apoptosis network.  In these
calculations, each gene has the same non-linear behavior given
by Eq. (2).  We found that the generic behavior was similar to that
presented in Fig. 2.  One example, where we used the
 Michaelis-Menten limit of Eq. (2), is presented in Fig. 5.  The statistics of the
death node (Fig. 5a) remain close to a normal distribution, while the
statistics of NFkB remains close to log-normal.  The geometry of
the network thus controls the signaling statistics even
in the presence of non-linearity.
\begin{figure}
\mbox{\subfigure[]{
  \epsfig{file=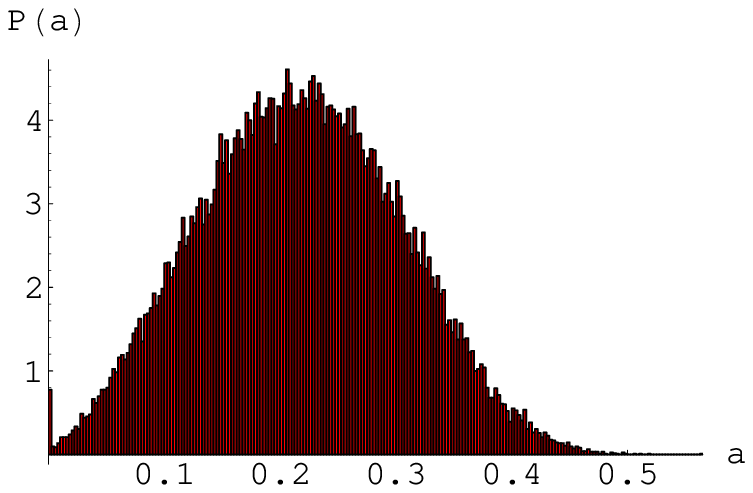 , width=0.8\columnwidth}}}\\
\mbox{\subfigure[]{
  \epsfig{file=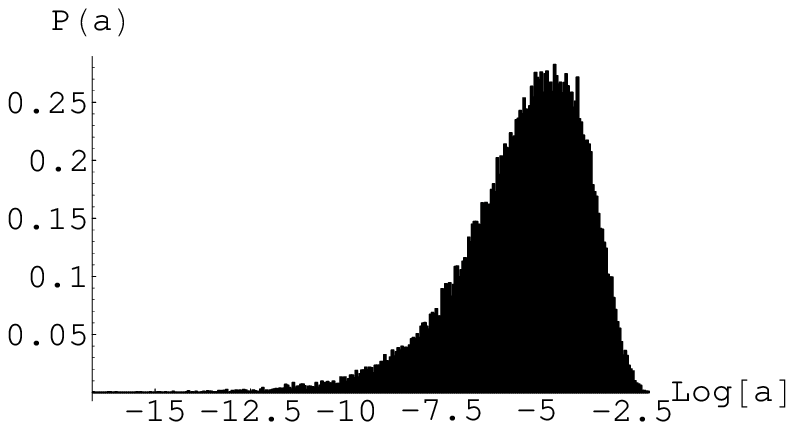 , width=0.8\columnwidth}
}}
\caption{Majority rule signaling with non-linearity.
The distribution of signals in heterogeneous cell
populations with a population size of 50,000.  These
distribution are for majority rule signaling with
positive, continuous gene activities calculated from the
signal by using the Michaelis-Menten law $a = s/(1+s)$. a) Signal
statistics at the output node is similar to a normal
distribution; b)
Signal statistics at the gene NFkB remains
close to log-normal despite the non-linear dependence of the
activity on the signal.  }
\end{figure}

An important feature absent from the Kegg apoptosis network is feedback.
The heavy dashed connection between CASP3 and CASP8 (nodes 42 and 28)
produces feedback which has recently been found to be important 
in the apoptosome\cite{eissing2005,bagci2006}.  This link leads
to feedback as illustrated in the subgraph of Fig. 6.  To elucidate 
the effect of feedback on signaling in the apoptosis network, we studied the
response of the network in Fig. 6 with random weights on the edges and a
range of signal strengths arriving at CASP8 ($a_1$ in Fig. 6).
\begin{figure}
{\centering \resizebox*{.7\columnwidth}{!}
{\rotatebox{0}{\includegraphics{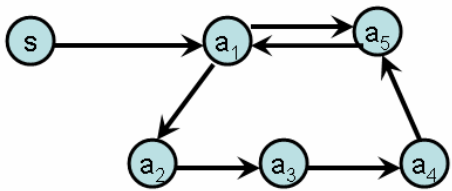}}}\par}
\caption{Feedback loop in the apoptosis network, following ref. \cite{bagci2006} }
\end{figure}
The equations for the signals, $s_i(t+1)$, arriving
at time $t+1$ at the five genes in Fig. 6 are,
\begin{eqnarray}
s_1(t+1) & = & S + w_{51} a_5(t) \\
s_2(t+1) & = &  w_{12} a_1(t) \\
s_3(t+1) & = &  w_{23} a_2(t) \\
s_4(t+1) & = &  w_{34} a_3(t) \\
s_5(t+1) & = &  w_{15} a_1(t) + w_{45} a_4(t).
\end{eqnarray}
In this equation $S$ is the input signal and
$w_{ij}$ is the strength of the signaling between
nodes $i$ and $j$.  In heterogeneous population
studies these links are taken to be random.
The activity of each gene at time $t+1$
is found using a non-linear relation to the
signal $s_i(t+1)$, given by the Hill equation,
\begin{equation}
a_i(t+1) = {c (s_i(t+1))^b)\over (1+d (s_i(t+1))^b)},
\end{equation}
where $b,c,d$ are model parameters.
The typical activities of the five genes, $a_i(t)$, as a function of the
input signal strength, $S$, are presented in Fig. 7 for three
types of Hill equation parameters, linear (top figure),
Michaelis-Menten (middle figure) and co-operative (bottom figure).
As observed in modeling using ODE's\cite{bagci2006}, co-operative signaling leads
to new behavior and a sharp onset of a transition between
a low activity state and a high activity state. The behavior of Fig. 7c
is typical of the co-operative case, and the location of
the jump discontinuity and the values of the gene activities
depend on the parameter values used in the simulations.  One
example is presented in this figure. We found that for
each parameter set there is a steady state response at long times
and this is the value that is plotted in the figures.
\begin{figure}
\mbox{\subfigure[]{
  \epsfig{file=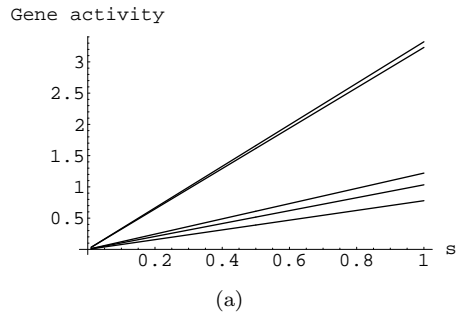 , width=0.7\columnwidth}}}\\
\mbox{\subfigure[]{
  \epsfig{file=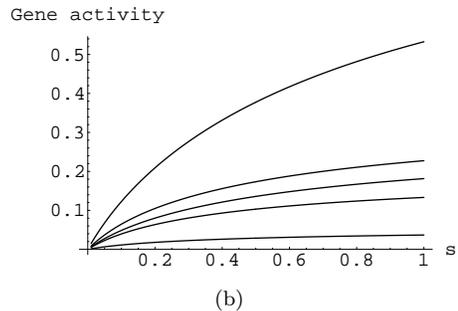 , width=0.7\columnwidth}}}\\
\mbox{\subfigure[]{
  \epsfig{file=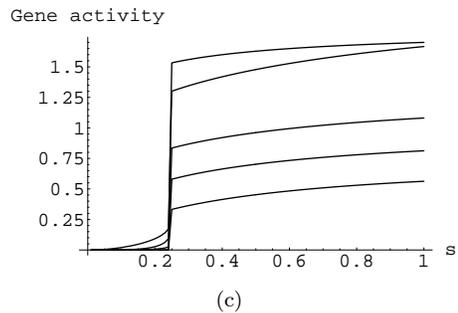 , width=0.7\columnwidth}}}
\caption{The steady state activity of genes in the feedback
loop illustrated in Fig. 6.  The top figure is the case
of linear signaling ($d=0,c=b=1$ in Eq. (10),
the middle figure is for the Michaelis-Menten case
($b=c=d=1$ in Eq. (10), and the bottom figure was found using Eq. (10) with
$c=2$, $d=1$ and $b=2$ corresponding to co-operative non-linearity.
In the top figure the link weights are $\omega_{12}=0.367$,
$\omega_{23}=0.846$, $\omega_{34}=0.754$, $\omega_{45}=0.617$,
$\omega_{51}=0.718$, $\omega_{51}=0.828$.  In the middle and bottom
figures the weights used were $(0.417,0.847,0.287,0.456,0.614,0.521)$ and
$(1.812,1,207,0.971,1.158,1.924,1.489)$ respectively.}
\end{figure}

We have also studied the effect of feedback on signaling statistics
in heterogeneous cell populations using the full apoptosis network,
Fig. 1.  Signaling in this network is carried out by using the OSA
procedure in combination with full iteration of the loopy subgraph
of Fig. 6.  In the linear and Michaelis-Menten cases
illustrated in the top two figures of Fig. 7,
feedback amplifies the signal, but does not qualitatively
change the statistics of death signals.  However in the case
of co-operative non-linearity where bistability and
strong sensitivity to the input signal occurs (see Fig. 7c), the
statistics of the death signal can become bimodal,
as illustrated in Fig. 8.  The signaling statistics can
then be controled by controling the feedback and non-linearity
in the network.
\begin{figure}
{\centering \resizebox*{\columnwidth}{!}
{\rotatebox{0}{\includegraphics{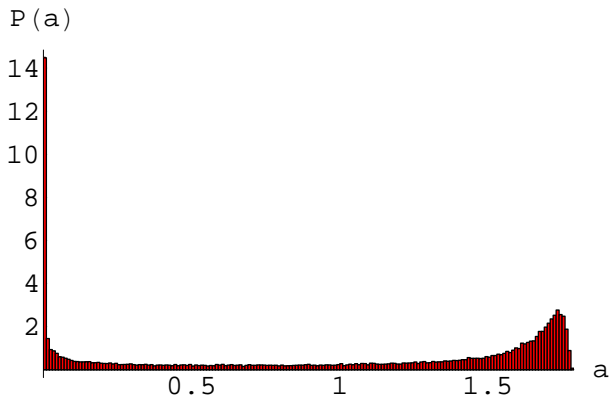}}}\par}
\caption{Death statistics in a heterogeneous population of 50,000 cells with
co-operative non-linearity and feedback.  The non-linearity parameters used
in Eq. (10) were $c=1.8, d=1, b=2$ and all of the link
weights were chosen to be random on the
interval $[0,c]$.}
\end{figure}

\subsubsection{Discrete models}

Discrete models enable
exhaustive search over the activity states of the genes and,
as elucidated in Section III, identification of the optimal
combinations for selective control.
In these models we discretize the weights and the activities into the
same number of discrete levels, so that $m_i = 0,1,...M$ and
 $|\omega_{ij}| = 1,...M$,
for a model where the activity level of a gene has a maximum
integer value of $M$.
We find that the generic
behaviors for large values of $M$ are similar
to that of the continuous models of the last subsection.
On the other hand, for binary networks
where each gene has activity zero or one, the behavior is quite different,
with a key novelty the fact that many genes have zero incoming signal  and
hence a decision must be made about their activity in this case.  Using the
momentum rule, where the state is maintained unless changed by an incoming signal,
leads to a strong dependence on initial conditions as the initial state is
unchanged unless a signal is received to change it.
As in Eq. (1), the signal, $s_l$ arriving at gene $l$ is a linear
superposition given by,
\begin{equation}
s_l = \sum_{j \epsilon n(l)} \omega_{jl} m_j,
\end{equation}
where $n(l)$ is the set of genes which signal directly
to gene $l$.   In the discrete models, the signal, $s_l$,
produced by the superposition rule above is
then normalised  to $s_l/n(l)$ where $n(l)$ is the number
of neighbors of gene $l$.  We use
several different linear and non-linear relations to
find the discrete activity of a gene, $m_l$, from the
normalized signal $s_l/n(l)$, as described below.
In all cases, if the signal arriving at a gene is negative,
the gene is completely inhibited and $m_l=0$, which
is a basic non-linearity in both continuous and
discrete signaling models.

In the linear rule, the discrete activity is found
from the normalized signal using,
\begin{equation}
m_l = \lceil c_1 {s_l\over n(l) M} \rceil~,
\end{equation}
where $c_1$ is a constant which we usually take to be $c_1=1$.
$\lceil x \rceil$ is the ceiling function, which
raises a floating point number, $x$, to its next largest integer
value.
For the logarithmic rule the discrete activity
is given by,
\begin{equation}
m_l=\lceil c_2 \ln \left(s_l/n(l)\right) \rceil~,
\end{equation}
where the constant $c_2$ is chosen so 
that the maximum signal corresponds to $m_l=M$. In our case with $M=10$,
we take $c_2=2.17$.  The sigmoidal rule is given by,
\begin{equation}
m_l=\lceil\frac{M}{\exp[-(s_l/n(l)-\alpha)/\beta]+1} -0.5\rceil~
\end{equation}
with the parameters $\alpha=20$ and $\beta=10$.
This has a form similar to the Fermi function in physics and to
dose-response curves in radiation therapy.
For $M=10$ the maximum normalized signal which
can arrive at a gene is $100$.  The functions (12-14)
are constructed so that for small signals
the gene activity is one, while signals of
maximum value yield gene activity $10$, ensuring that
all possible relations between signal and activity are
represented.  These behaviors are presented in Fig. 9a.
\begin{figure}
\mbox{\subfigure[]{
  \epsfig{file=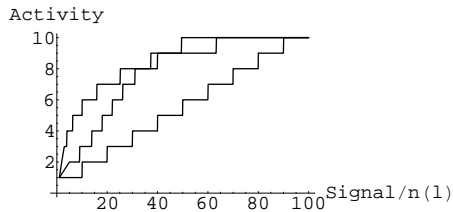 , width=0.7\columnwidth}}}\\
\mbox{\subfigure[]{
  \epsfig{file=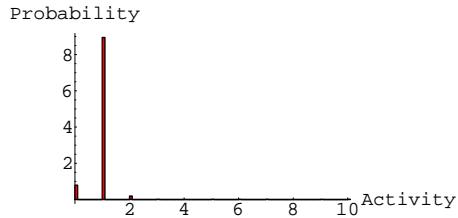 , width=0.7\columnwidth}}}\\
\mbox{\subfigure[]{
  \epsfig{file=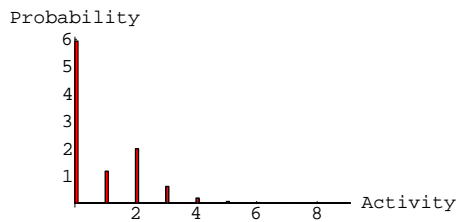 , width=0.7\columnwidth}}}\\
\mbox{\subfigure[]{
  \epsfig{file=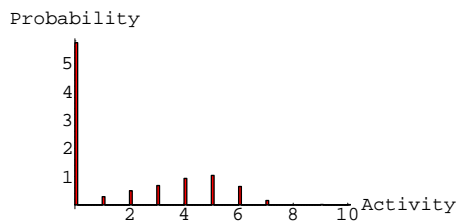 , width=0.7\columnwidth}}}
\caption{a) Relations between the normalized signal, $s_l/n(l)$,
arriving at a gene and it's discrete activity, from the top the
behaviors are for the logarithmic function, Eq. (13), the sigmoidal function,
Eq. (14) and the linear function, Eq. (12).
The death statistics for
these three models: b) linear function;
c) sigmoidal function; d) logarithmic function.
It is clear that the distribution of death activities becomes bimodal
and broader as we go from the linear function to the logarithmic function.}
\end{figure}

The death statistics resulting from these discrete models are presented in Figs.
9b-d.  The discrete linear model leads to unimodal statistics, however the
sigmoidal and logarithmic functions lead to bimodal statistics.  This is due
to the fact that the latter functions amplify small signals, as is evident in Fig.
9a.  In the next section we use these three discrete models in studies of
selective control.

\subsection{Statistics of selective control}
Selective control aims at changing the life/death signal of one member of a
population with a minimal change of the remaining members of the population. We
address in this section the question of how network topology and general signal
propagation properties can be used to design strategies for selective control.
The control of the life/death signal is realized by acting with external
perturbations (drugs) on the nodes and on the signaling flow. As we have seen
in the previous sections, the characteristics of the signaling through the
network can be strongly dependent on the OSA rule. We will see below how
strategies for selectivity are affected by these rules. General strategies for
selectivity can be inferred by analyzing the statistics of the nodes involved
in selective perturbations. For a given network topology, we can identify the
nodes that are more likely to appear in a selective perturbation, and analyze
their correlations. Important insight, such as the role of balancing
pro-apoptosis and anti-apoptosis perturbations in selectivity, can be obtained
from this analysis. Moreover, the correlation analysis revealed that for robust 
populations selective nodes tend to be the ones that produce the stronger change
in the output signal. The opposite is true in the case of weak populations, for
which selectivity is improved by acting on nodes that
produce weak signal changes in the output.

\subsubsection{Exhaustive search in discrete models}
In this subsection we carry out an exhaustive search of selective perturbations by
discretizing the control parameters and signaling variables.
In the next section, we will show how some of the key features of
selectivity statistics can be captured with a simplified method based on
linear programming optimization, which is less demanding from a computational
point of view.

We start by generating a population of different apoptosis networks with the
same topology, but random values for the initial gene expressions $m^0_i
(t=0)\in [0,M]$ and random strength of the links $\omega_{ij}\in[-M,-1]$ for
inhibition, and $\omega_{ij}\in [1,M]$ for stimulation ($M=10$ in the numerical
calculations).  The chosen population needs to represent living cells, {\it
i.e.} it must have the property
\begin{equation}
s_{o,\lambda}<\bar s_o; ~~~~~\forall \lambda, \label{eq:living}
\end{equation}
where $\bar s_o$ is the signal life/death threshold value, $ \lambda$ is the
index labeling individuals in the population, and $s_{o,\lambda}$ is the output
node signal. We take homeostasis into account by adding the constraint that
each individual $\lambda$ remains alive under fluctuations on the input nodes.
After $s_{o,\lambda}$ has been calculated for a given input, we let the input
nodes fluctuate and we recalculate its value. If the output of a network is less
than a threshold value, that is
$s_{o,\lambda} < \bar s_o $, for ten random fluctuations of the input nodes,
then we keep the individual $\lambda$ in the living population.  In the 
analysis of selectivity, described below, a population 
of 100 was chosen as the number of different cell types in the human 
body is of this order.  Preliminary studies indicate that, as expected,  
selective control is easier for 
smaller populations, provided the number of control nodes is fixed.

Once we have created a living population, we can start to study the effect of external
perturbations on the nodes. These perturbations represent the effect of drugs
that stimulate or inhibit one or more gene. We will therefore
represent the effect of the drug by changing the gene expression levels in the
nodes by $\delta m_i$.  We will say that an individual
$\bar\lambda$ can be {\it selectively controlled} if we can find a perturbation
on the gene expression levels $\delta m_i$ with the property
\begin{eqnarray}
s_{o ,\bar\lambda}>\bar s_o;~~~~~~~~~~~~~~~ \\
s_{o,\lambda}<\bar s_o; ~~~~~\forall \lambda \ne \bar\lambda~.
\end{eqnarray}
Though there are 48 nodes, 11 are input nodes and one is the output node, so 
there are 36 control nodes in the network.  There are 
$M-1$ possible perturbations on each control node so the total number of
possible perturbations on all the nodes, $(M-1)^{36}$, which is too large to be explored
exhaustively. Therefore, we considered 
perturbations that act only on $k$-subsets of all possible perturbations. For
$k=1$ this means that we are considering only single node perturbations, which
requires a search over $36 (M-1)$ possibilities. For $k=2$ we are considering pairs of
nodes, with $630(M-1)^2$ perturbation combinations. For an arbitrary
$k$-subset the number of combinations is $C^{36}_k (M-1)^k$.

In Table \ref{tab:sel} we present the selectivity results for ten different
populations with 100 individuals each and three different OSA rules. 
The columns the Table II represent different $k$-subsets for the three different rules. The
value for the threshold $\bar s_o$ was set to 1. In the linear case, the
average percentage of individuals that can be selectively killed within the
$k=3$ subset is about 63\%. We have found that this average depends strongly on
the value of the threshold, and it decreases for higher threshold values. For
instance, by setting the threshold at $\bar s_o=7$ the average selectivity
(within the $k=3$ manifold) is reduced to 8.3\%. The logarithmic and sigmoid
rules give a overall higher selectivity, with an average selectivity of 63.5 \%
and 73.6 \%,  respectively, for death threshold $\bar s_o =1$.
\begin{table}
\begin{tabular}{||c|c|c||c|c|c||c|c|c||}
\multicolumn{3}{c}{Linear}& \multicolumn{3}{c}{Sigmoid}&
\multicolumn{3}{c}{Logarithmic}\\ \hline
 k=1& k=2 & k=3 &k=1&k=2&k=3& k=1& k=2 &k=3 \\ \hline
 4 & 22& 58 &1 &32 & 76         & 9&30 &64\\
 2 & 25& 61&4 &35 &71           & 9 &34 &62 \\
 2&21&67& 4& 31& 77             & 9&28 &64 \\
 3&36&68 &1 & 37& 66            &10 &29 &58 \\
 2&30 &68 &3 &28 &74            & 8&32 &63\\
 2&39 & 71 & 2& 33& 74          & 9& 34&62\\
 1&26&53& 3& 37&76              &4 &30 &68\\
 1&28&60& 5& 31& 66             &10 &33 & 63\\
 1&28 &66 &3 &37 &76            &4 &30 &68 \\
 1&23&57& 5& 37&80              &10 &33 &63
\end{tabular}
\caption{Number of individuals that can be selectively controlled in 10
populations with 100 individuals each. The different columns refer to
different $k$-subsets and different OSA rules, and with the death threshold 
$\bar s_o=1$.} \label{tab:sel}
\end{table}
\begin{figure}
{\centering \resizebox*{\columnwidth}{!}
{\rotatebox{0}{\includegraphics[width=6.5 cm]{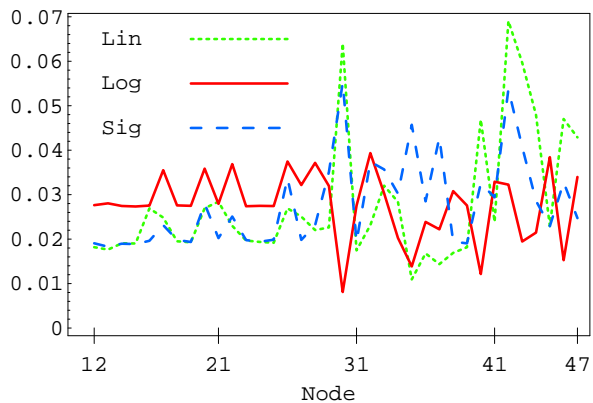}}}\par}
\caption{(Color online) Distribution of nodes entering in selective
combinations for the Linear (dotted), Sigmoid (dashed), and Logarithmic (solid)
OSA rules.} \label{fig:freq}
\end{figure}

We have studied the statistics of nodes entering in selective combinations. In
the linear case, single-node selectivity is obtained by acting on BAD
(30), IkBa (33), NFkB (34), BAX (40), CASP3 (42), CASP7 (43) or TP53 (44).
These nodes can be divided into those that are pro- and anti-apoptosis, based
on their average effect on the output signal. Single drug selective apoptosis
is generally induced by stimulating a pro-apoptosis node, or by inhibiting an
anti-apoptosis node. For instance, nodes IkBa (33) and NFkB (34) have on
average an anti-apoptosis behavior, so they are mainly associated with negative
$\delta m$. The dotted line in Fig.~\ref{fig:freq} shows the distribution of
nodes entering in all the selective combinations found using the linear OSA
rule. Notice the peaks corresponding to the same nodes that can induce
single-drug selectivity discussed above. In the figure, we also plot the
distribution of nodes for the sigmoid (dashed line) and the logarithmic (solid
line) rules. The sigmoid and the linear rule identify a very similar set of
nodes that are more likely to enter in selective perturbations. Both BAD (30)
and CASP3 (42) are strongly present in the linear and sigmoid model. The
sigmoid rule slightly enhances the number of combinations in which the
inhibition of anti-apoptosis genes represents the mechanism for selectivity,
see e.g. CHUCK (32), IkBa (33), NFkB (34), BIRC2 (35) and BCL2 (37). Both
linear and sigmoid rules suggest that selectivity might be better achieved by a
direct up-regulation of caspases, or by acting on pro- or anti- apoptosis
proteins of the BCL2 family. The logarithmic rule results
in a different distribution, suggesting a different strategy for selectivity.
Most of the selective combinations in the logarithmic case involve nodes
upstream in the signaling network, indicating that the best strategy for
selectivity is an action on cell-membrane FAS or TNF pathways.

\begin{figure}
{\centering \resizebox*{\columnwidth}{!}
{\rotatebox{0}{\includegraphics[width=6.5 cm]{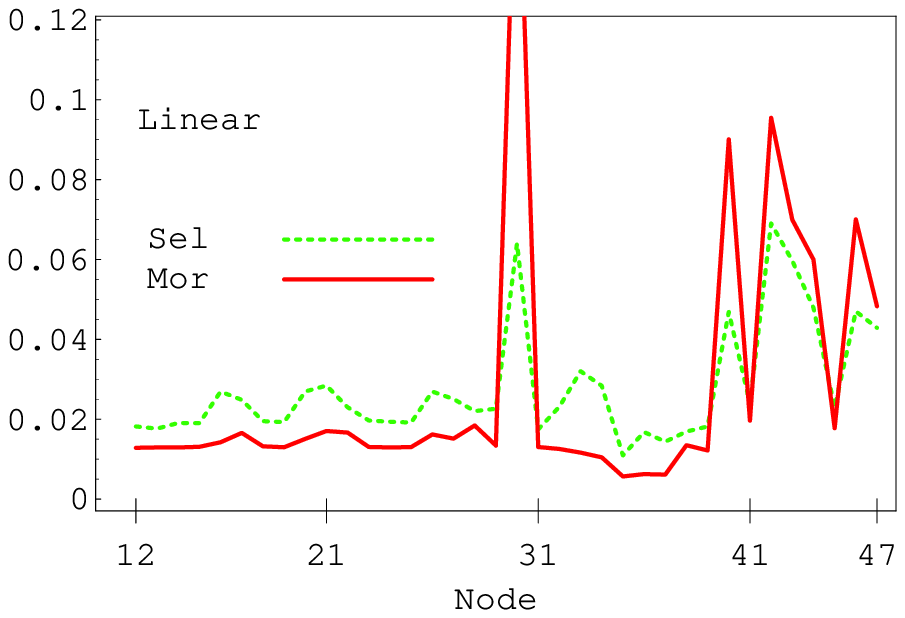}}}
\par}
{\centering \resizebox*{\columnwidth}{!}
{\rotatebox{0}{\includegraphics[width=6.5 cm]{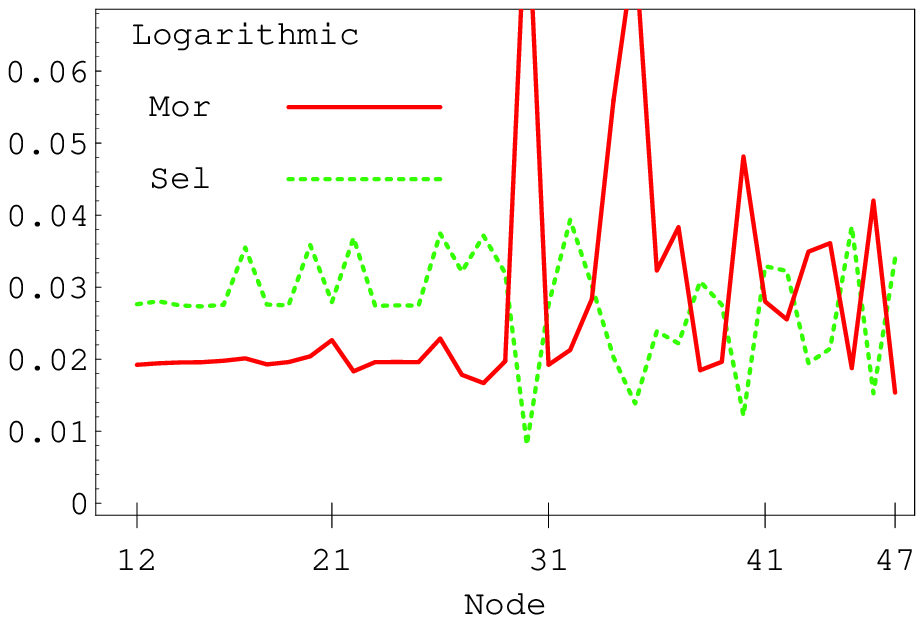}}}
\par}
\caption{Distribution of nodes appearing is selective and mortal perturbations
in the linear (top panel) and logarithmic (lower panel) OSA rules. The
life/death threshold is set to $\bar s_o=1$ for the two rules. The correlation
between the mortality and selectivity distributions is 0.88 for the linear case
and -.80 for logarithmic case.} \label{fig:corrmorsel}
\end{figure}

Interesting correlations can be observed between the statistics of selective
and mortal perturbations. Any perturbation kills at least
one member of the population is defined as {\it mortal}. Selective
perturbations are mortal perturbations, but there are many more nonselective
mortal combinations which kill more than one individual in the population.
Fig.~\ref{fig:corrmorsel} show the distribution of nodes entering in selective
and mortal combinations in the case of linear (top panel) and logarithmic OSA
rules (lower panel). The linear rule shows a strong correlation between nodes
that appear in selective combinations and nodes that appear in mortal
combinations. The correlation between the mortality and and selectivity
distribution is 0.88. The logarithmic case is completely different
(lower panel) and exhibits a strong anti-correlation between selectivity
and mortality (-0.80). The anti-correlation in the logarithmic case can be
reduced by increasing the threshold, and switches to correlation for values of
the threshold larger than the average value of the output signal. This behavior
can be understood in the following way. For a very high life/death threshold it
is very difficult to find individuals that can be killed. In that case we can
say that the population is very robust with respect to external perturbations.
Therefore, if one individual can be killed, the perturbation that kills that
individual is very likely to be selective. We have observed that by pushing the
threshold to very high values the correlation between selectivity and mortality
approaches one. The nodes that are involved in this case are the ones that are
able to produce the strongest change in the output, and are the same nodes
usually involved in many mortal combinations. On the other hand, if the
population is weak, nodes that are highly mortal are likely to kill more than
one individual at the same time, therefore selectivity is associated with nodes
that are less mortal, i.e. those that produce small changes in the output signal. The
robustness or weakness of a population is determined not only by the value of
the life/death threshold, but also by the OSA rules giving different signaling
statistics as discussed in the previous section. In fact, the behavior in
Fig.~\ref{fig:corrmorsel} was obtained using the same threshold ($\bar s_o=1$).
There, the correlation/anticorrelation is a consequence of the higher
sensitivity of the logarithmic rule to external perturbations, which implies
that the population is much weaker compared to the linear case.

\begin{figure}
{\centering \resizebox*{\columnwidth}{!}
{\rotatebox{0}{\includegraphics[width=6.5 cm]{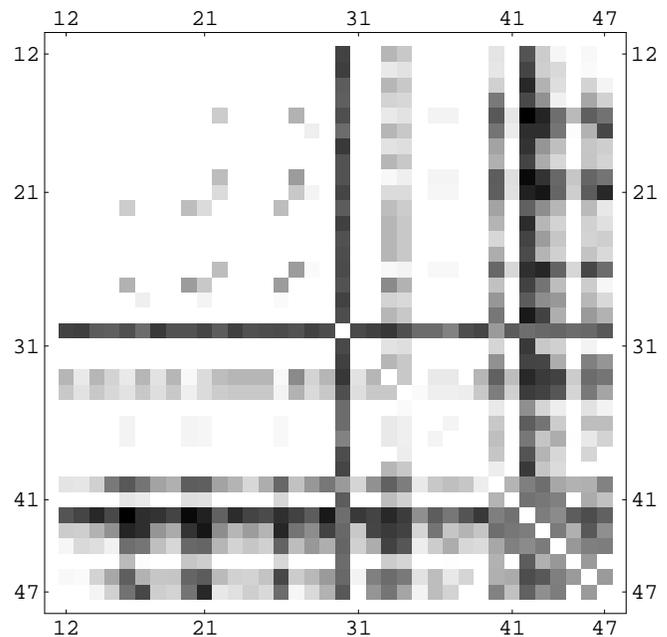}}}
\par}
\caption{Correlation matrix for nodes appearing in the same selective
perturbation in the linear OSA rule. Darker dots indicate that the two nodes
given by the row and column of the matrix appear more often in the selective
drug combinations.} \label{fig:corr}
\end{figure}

\begin{figure}
{\centering \resizebox*{\columnwidth}{!}
{\rotatebox{0}{\includegraphics[width=6.5 cm]{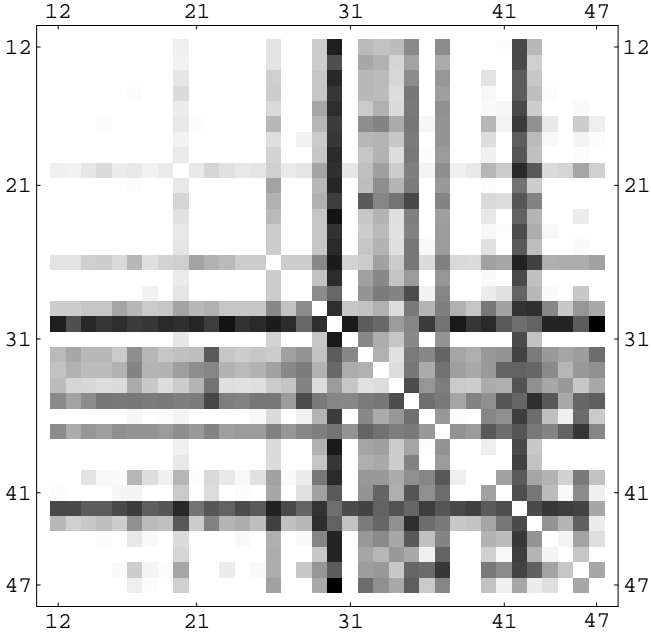}}}
\par}
\caption{Correlation matrix for nodes appearing in the same selective
perturbation in the sigmoidal OSA rule. Darker dots indicate that the two nodes
given by the row and column of the matrix appear more often in the selective
drug combinations.} \label{fig:corrsig}
\end{figure}

\begin{figure}
{\centering \resizebox*{\columnwidth}{!}
{\rotatebox{0}{\includegraphics[width=6. cm]{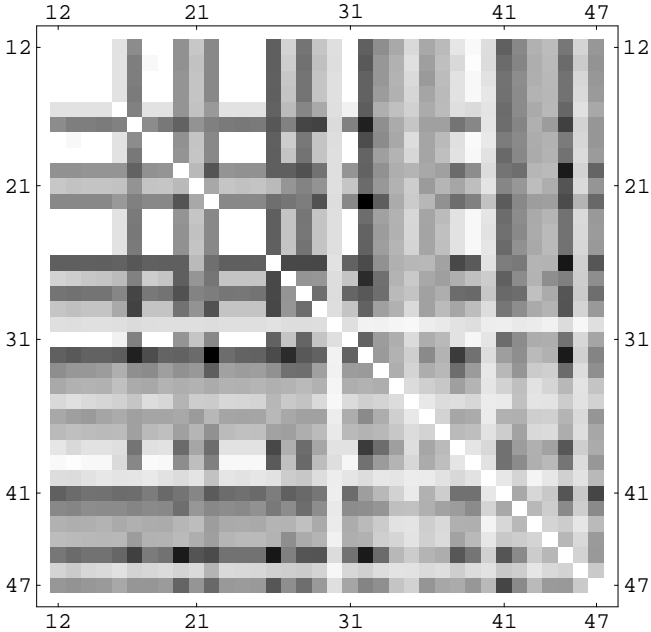}}}\par}
\caption{Correlation matrix for nodes appearing in the same selective
perturbation in the logarithmic OSA rule. Darker dots indicate that the two
nodes given by the row and column of the matrix appear more often in the
selective drug combinations.} \label{fig:corrlog}
\end{figure}

We have analyzed correlations between two nodes appearing in selective
combinations. For the linear rule, this is shown in Fig.~\ref{fig:corr} using a
correlation matrix plot. Darker dots indicate a higher number of selective
combinations containing the two nodes given by the row and column of the
matrix. Notice the strong presence of the mitochondrial BAD (30) and BAX (40),
and the caspases CASP3 (42) and CASP7(43). However, these nodes often appear in
combination with other nodes, many of which have an anti-apoptosis character.
The balance of pro and anti-apoptosis perturbations is the key element which
increases the selectivity from 1.5\% in the $k=1$ case to the 63\%  in the
three-node perturbation. Notice also that in this linear case the nodes
involved in selective combinations are often downstream (i.e. close to the
output node) in the signaling network.
We show in Fig.~~\ref{fig:corrsig} and Fig.~~\ref{fig:corrlog} the same
correlation matrices in the case of the sigmoid and logarithmic OSA rules. In
these models more nodes are involved in the selective combinations. The
correlation pattern for the sigmoid rule shows strong similarities to the
linear case. However, the logarithmic rule results in a qualitatively different
pattern in the correlation matrix. Notice for instance that the role of BAD
(30), which was dominant in the linear and sigmoid rule, is strongly reduced in
this case. In contrast to the linear and sigmoid case, nodes in long pathways
dominate in the selective combinations.
Overall, the presence of nonlinearity
in the OSA rules seems to enhance the possibility of selective control. This is
also confirmed by the trend in the total number of $k=3$ selective
combinations, being $9\times 10^6$, $11\times 10^6$, and $26\times 10^6$  for the linear, sigmoid
and logarithmic OSA rules, respectively.

\subsubsection{Linear programming methods}
The exhaustive search method, discussed in the previous subsection, for
selective combinations becomes very demanding when combinations with a large
number of drugs are involved. A different approach consists in defining an
effective model for the dependence of the output signal on the perturbations.
The effective model is derived from the original OSA approach using a
sensitivity analysis, and some of the selective perturbations found with the
effective model are also selective perturbations for the original OSA problem.
The advantage of the effective model is that selective combinations can be
efficiently obtained by linear programming methods~\cite{cormen2001}. We will
analyze below the statistics of the selective combinations for the linear OSA
method obtained through the effective model. The form of the solutions in many
cases involves a number of nodes larger than three, in contrast to the
exhaustive method discussed in the previous section. The effective method
therefore provides a different sampling of the full space of selective
combinations. However, we will see that this different sampling leads to very a
similar statistics for the selective nodes.

\begin{figure}
{\centering \resizebox*{\columnwidth}{!}
{\rotatebox{0}{\includegraphics[width=6.5 cm]{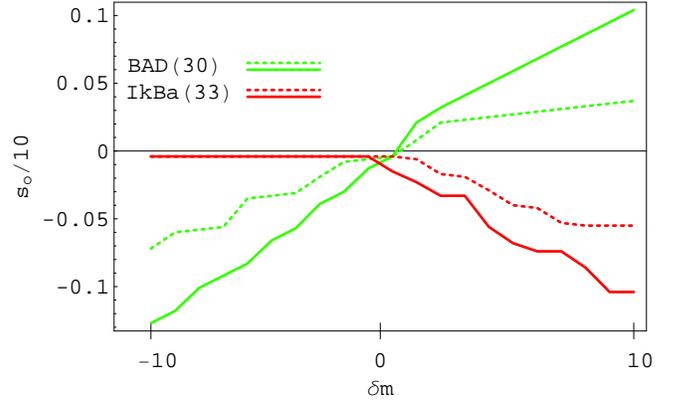}}}\par}
\caption{(Color online) Output signal as a function of the perturbation
strength on a pro-apoptosis (BAD, red increasing line) and anti-apoptosis node
(IkBa, green decreasing line). The dashed line and the solid line refers to two
different individuals.} \label{fig:proanti}
\end{figure}
\begin{figure}
{\centering \resizebox*{\columnwidth}{!}
{\rotatebox{0}{\includegraphics[width=6.5 cm]{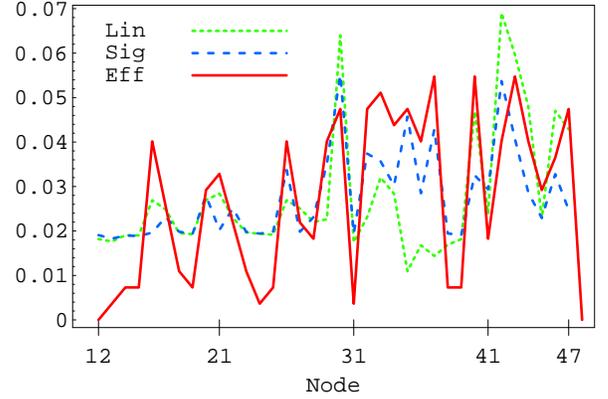}}}\par}
\caption{ (Color online). Distribution of nodes entering in selective
combinations obtained with the Linear OSA (dashed) and with the effective
linear programming method (solid line) } \label{fig:comp}
\end{figure}

The output signal from a OSA can be written as
\begin{equation}
s_{o,\lambda}(\delta m_{12},\delta m_{13}, \dots \delta m_{47})~,
\end{equation}
and depends in a nonlinear way on the perturbations of the internal nodes. A
typical single node dependence is shown in Fig.~\ref{fig:proanti} for a
pro-apoptosis BAD (30) and an anti-apoptosis IkBa (33) node for two different
individuals (solid and dashed lines) as a function of the strength of the
single node perturbation. Notice that the single drug dependence follows a
step-like dependence as a consequence of the discreteness of the gene
expression levels. Moreover, due to the constraint that the activity
must remain positive, there are regions in which the dependence on the external
perturbation saturates. The effective linear model can be defined as
\begin{equation}
s_{o,\lambda}(\delta m_{12},\delta m_{13}, \dots \delta m_{47})\sim
s_{o,\lambda}(0)+\sum_{i=12}^{47}c_{\lambda,i}{\delta m_i}~
\end{equation}
where the coefficient $c_{\lambda, i}$ can be estimated by approximating the
single drug dependence such as the ones in Fig.~\ref{fig:proanti} with a linear
dependence using interpolation methods. The linear interpolation works well
only for some nodes and individuals, since in many cases the dependence is
non-monotonic and highly nonlinear. Moreover, notice that we are neglecting
higher order many-node effects that are included in the OSA approach. However,
we will see below that this does not affect considerably the statistics of the
genes that are more likely to appear in selective combinations. The
interpolation method also allows us to identify nodes that lead to the highest
variations of the output, and the selective combinations can be restricted to
nodes within that set. The smallest coefficients $c_{\lambda,i}$ can therefore
be neglected. The reduction of the control parameter phase space by sensitivity
analysis is often a key element in global optimization problems. This was shown
explicitly in the case of parameter identification in biochemical reaction
networks~\cite{feng2004a,feng2004b}. Once the coefficient $c_{\lambda, i}$ and
the threshold value $\bar{s}_o$ have been fixed, the selectivity problem can be
recast in the form of a linear programming optimization problem where we
minimize the cost function
\begin{equation}
J(\delta m_{12}, \dots, \delta m_{47})=\sum_{i=12}^{47}|\delta m_i|
\label{eq:min}
\end{equation}
on the polytope defined by the constraint equations
\begin{eqnarray}
\sum_{i=12}^{47}c_{\lambda,i}{\delta m_i}<\bar{s}_o\\
\sum_{i=12}^{47}c_{\bar{\lambda},i}{\delta m_i}>\bar{s}_o~.
\end{eqnarray}
The solution provided by the linear programming method is optimal in the sense
that it gives the global minimum of the function in Eq.~(\ref{eq:min})~\cite{cormen2001}.

We show in Fig.~\ref{fig:comp} the statistics of the nodes found in the linear
programming optimization (solid line) compared to the same statistics obtained
with the exhaustive search described in the previous section (dotted line for
linear, dashed line for sigmoid). The two approaches identify basically the
same nodes as the ones that are more likely to be present in selective
combinations. Notice the strong presence of BAD (30), BAX (40) and Caspases
(42-44) in the three approaches. Also, the relatively strong peaks at IL-3R
(17), PI3K (20), and AKT3 (26) are captured by the effective approach. These
peaks suggest the possibility of selective control by acting on the AKT
signaling pathway.

We also have used the optimal control parameter $\delta m_{i}^*$ obtained with
the linear programming on the original nonlinear OSA and checked the system for
selectivity. Typically, we found that the linear programming solution is only
partially selective for the original OSA rule.  The drug combinations found from the 
linear programming method can selectively kill only 20\% of the 100 individuals 
in the survivor population, which is considerably 
smaller than that found using exhaustive search.
However, many selective combinations found by the linear programming
approach are {\it quasi-selective}, in the sense that
they kill two/three individuals in the population rather than one which 
is requirement for selectivity. This quasi-selectivity
captured by the linear programming method is at the origin of the strong
similarity in the distributions of Fig.~\ref{fig:comp}.

\section{Discussion}

Analysis of the statistical behavior of genes in the apoptosis network
illustrates the importance of local connectivity on gene activity variations.
Genes which receive signals from many parallel paths exhibit normal statistics
while genes while lie at the end of a single pathway are prone to large
statistical variations and highly skew statistics.  Though non-linearity alone
does not alter these broad conclusions, the combination of non-linearity and
feedback may lead to bimodal statistics.  This bimodality occurs due to
combination of feedback and bistability\cite{bagci2006}, which in our signaling
models is reflected in a sharp rise in activity with signal strength.

The concept of selective control in heterogeneous cell populations was
developed,  using the apoptosis network as an illustrative example. Selective
control within a heterogeneous cell population is the ability to control one
member of the population while leaving the other members relatively unaffected.
General selective control strategies that are only dependent on the topology of
the network and signaling rules can be inferred from analysis of networks with
random link weights. For instance, linear and sigmoid rules identify the same
set of nodes that are most efficient in selective control. These nodes can
lead to a high degree of selectivity within a given population, especially by
balancing pro- and anti- apoptosis perturbations. We have explored two methods
for the study of selectivity. The first is an exhaustive search method limited
to three node perturbations. The second is an effective linear model, based on
interpolation of single node sensitivity, in which the selective combinations
can be found by linear programming optimization. The two approaches identify
the same strategies for selectivity. We have also identified a general rule
that relates the life/death switching robustness of a population to the optimal
selectivity strategy. Selectivity is promoted by acting on the least
sensitive nodes in the case of weak populations, while selective control of
robust populations is optimized through perturbations of more sensitive
nodes. More generally selective control is a computational challenge in a broad
range of systems biology problems where intervention needs to be directed at
subsets of a diverse population.

At a more practical level, high throughput experiments with heterogeneous cell
lines could be designed in such a way that the selectivity optimization process
is part of a closed-loop control system~\cite{ku2004}. Single drug measurements
could be used to obtain directly the sensitivity in a given heterogeneous cell
population. The linear programming optimization method can then use the
sensitivity measurements to identify selective combinations in a model free way. The
sensitivity/selectivity optimization can then be improved through iterated experiments
in which various computational algorithms and experimental protocols are
integrated.

\section{Methods}

Two programs were used to generate the signaling statistics and to 
analyse selectivity.  One procedure was written in Mathematica and the 
other in C++.  These independent programs were used to check the 
accuracy of the numerical analysis of the signaling procedures and 
the signaling statistics.  

The software used to run the exhaustive search tests is written in C++ and compiled for both the Linux 32 bit and 64 bit operating systems. 
We used two clusters, available at the Burnham Institute for Medical Research, Falcon and Bsrc. Both of them are Portable Batch System (PBS) clusters, so that they have a native implementation of queues management enabling submission of the same program multiple times to achieve virtual parallelism. Falcon has 128 CPUs organized in 64 nodes and each of them is a x86 32 bit 2.4 GHz with 1 GB of private memory; Bsrc has 128 CPU organized in 32 nodes and each of them is a x86 64 bit 2.0 GHz with 2 GB of private memory.
Exhaustive search requires an exponential computational time in the number of nodes and projecting from the time needed for combinations of 1,2,3 nodes, that is respectively 1 minute, 
1.5 hours and 7.5 hours, we estimate that around 90 days would be needed to study all the combinations of 4 drugs. The calculations on the continuous models and the linear programming optimization were implemented using Mathematica on a personal computer. The NMinimize function in Mathematica finds the global minimum when the objective function and constraints are linear.

\section*{Acknowledgments}
PMD thanks Leon Glass and Bard Ermentrout for useful discussions.
\bibliography{signaling}

\end{document}